\documentclass[fleqn,twoside]{article}
\usepackage[headings]{espcrc2}

\readRCS
$Id: espcrc2.tex,v 1.2 2004/02/24 11:22:11 spepping Exp $
\usepackage{graphicx}
\usepackage[figuresright]{rotating}

\newcommand{\AmS}{{\protect\the\textfont2
  A\kern-.1667em\lower.5ex\hbox{M}\kern-.125emS}}

\newcommand{\hepph}[1]{{\tt hep-ph/#1}}
\newcommand{\NP}[1]{Nucl.\ Phys.\ {\bf #1}}
\newcommand{\PL}[1]{Phys.\ Lett.\ {\bf #1}}

\newcommand{\PR}[1]{Phys.\ Rev.\ {\bf #1}}

\newcommand{\PRL}[1]{Phys.\ Rev.\ Lett.\ {\bf #1}}

\newcommand{\eq}[1]{eq. (\ref{#1})}

\def\beq{\begin{equation}}
\def\eeq{\end{equation}}
\def\beqa{\begin{eqnarray}}
\def\eeqa{\end{eqnarray}}
\def\ifm{\ifmmode}
\def\msb{\ifm \overline{\rm MS}\,\, \else $\overline{\rm MS}\,\, $\fi}
\def\msbns{\ifm \overline{\rm MS}\, \else $\overline{\rm MS}\, $\fi}
\def\gtil{{\widetilde {G}}}

\renewcommand{\a}{\alpha}
\newcommand{\e}{\epsilon}

\renewcommand{\b}{\beta}

\title{Refining threshold resummations}

\author{Eric Laenen\address{NIKHEF Theory Group, Kruislaan 409, 
        1098 SJ Amsterdam, The Netherlands, and \\
        Institute for Theoretical Physics, Utrecht University,
        Leuvenlaan 4, 3584 CE Utrecht, The Netherlands}\thanks{Work 
        supported by the Foundation for Fundamental Research of
        Matter (FOM) and the National Organization for Scientific 
        Research (NWO).} and
        Lorenzo Magnea\address{Dipartimento di Fisica Teorica, 
        Universit{\`a} di Torino, and \\  INFN, Sezione di Torino, 
        Via P. Giuria 1, I-10125 Torino, Italy}\thanks{Work supported 
        in part by MIUR under contract $2004021808\_009$. E-mail:
        {\tt magnea@to.infn.it}} }

\runtitle{Refining threshold resummations}
\runauthor{E. Laenen and L. Magnea}

\begin{document}

\begin{abstract}
We describe some recent refinements of the techniques of threshold
resummation, with emphasis on the usefulness of dimensional
regularization when applied to nonabelian exponentiation. Threshold
resummation is now under theoretical control for DIS and electroweak
annihilation cross sections all the way to the fourth tower of
logarithms and up to corrections suppressed by powers of the threshold
variable.
\vspace{1pc}
\end{abstract}


\maketitle

\section{INTRODUCTION}

Soft and collinear gluon radiation in perturbative QCD is well-known
to have special properties of universality and factorization, which
are related to its semiclassical
nature~\cite{Low:1958sn,Burnett:1967km} and strongly tied to gauge
invariance~\cite{Collins:1989gx}. These properties are at the heart of
our understanding of strong interactions at high energies, as they
allow us to define infrared and collinear safe quantities to all
orders in perturbation theory, and lead to factorization of long
distance contributions in hadronic processes. Understanding soft and
collinear gluons is however not merely a theoretical exercise: it
has important phenomenological consequences. Even when dealing with
observables which are finite order by order in perturbation theory, one
still finds in many cases that the cancellation of long distance
singularities leaves behind finite but large contributions, typically
logarithms of large ratios of kinematic scales. These contributions
need to be resummed to all orders to have reliable predictions, in
some cases even to get just a qualitative agreement with experimental
data. Resummation is a well developed
technology~\cite{Sterman:1987aj,Catani:1989ne}, deeply connected to
the universality of soft and collinear singularities.

In perturbative QCD, the use of dimensional
regularization to regularize mass singularities is
not required in theory, but is in practice.
Here we would like to discuss
some recent developments concerning the resummation of threshold
logarithms, while emphasizing the simplicity and elegance which can be
achieved by making extensive use of dimensional regularization in
dealing with the underlying pattern of soft and collinear
divergences. In essence, resummation of threshold logarithms is always
a consequence of exponentiation of soft and collinear
singularities. Nonabelian exponentiation implies that double poles of
infrared and collinear origin can be organized in a predictive manner
in terms of an exponent containing only simple poles,
\beq
  \sum_k \alpha_s^k \sum_p^{2 k} c_{k p} \epsilon^{- p}
  \rightarrow \exp \Big[ \sum_k \alpha_s^k \sum_p^{k + 1} d_{k p}
  \epsilon^{- p} \Big] \, .
\label{sumpo}
\eeq
After the cancellation of singularities between real and virtual
contributions has taken place, each pole leaves behind a logarithm
$L$, with argument for example the Mellin variable in DIS, where $L =
\log N$.  A similar pattern of exponentiation then applies to these
logarithms, which are organized as
\beqa       
  && \hspace{-4mm} \sum_k \alpha_s^k \sum_p^{2 k} \hat{c}_{k p} L^p
  \rightarrow \exp \Big[ L \, g_1 (\alpha_s L) + g_2 (\alpha_s L) 
  \nonumber \\ & & \, + \,
  \alpha_s \,  g_3 (\alpha_s L) + \ldots \Big] \, .
\label{sumlo}
\eeqa
We take then the viewpoint that in order to resum threshold logarithms
we first need to resum infrared and collinear poles. Since all such
resummations involve the coupling evaluated at a soft scale, the first
tool must be a dimensionally regularized version of the running
coupling. As is well known, in $d = 4 - 2 \epsilon$, with $\epsilon <
0$ for infrared regularization, this coupling must satisfy the
equation
\beqa
  \mu \frac{\partial \overline{\alpha}}{\partial \mu} & \equiv &
  \b (\e, \overline{\alpha}) = - 2 \, \e  \,
  \overline{\alpha} + \hat{\b} (\overline{\alpha}) \, , 
  \nonumber \\
  \hat{\b} (\overline{\alpha}) & = & - \frac{\overline{\alpha}^2}{2
  \pi} \sum_{n = 0}^\infty b_n \left(
  \frac{\overline{\alpha}}{\pi} \right)^n \, ,
\label{dbeta}
\eeqa
where $\hat{\b} (\overline{\alpha})$ is the ordinary $4$-dimensional
$\beta$ function, with $b_0 = (11 C_a - 2 n_f)/3$. At one loop, the
solution is
\beq
  \overline{\alpha} \left(\mu^2 \right) = \frac{\alpha_s (\mu_0^2)}{
  \left[ \left(\frac{\mu^2}{\mu_0^2} \right)^\epsilon -
  \frac{b_0}{4 \pi \epsilon} \left( 1 - \left(\frac{\mu^2}{\mu_0^2}
  \right)^\epsilon \, \right) \alpha_s(\mu_0^2)
  \right]} ,
\label{alphas}
\eeq
which reduces to the well-known limit as $\epsilon \to 0$. Working
with \eq{dbeta} leads to compact resummed expressions for both
amplitudes and cross sections, which can be readily compared with
Feynman diagram calculations.  This was first done for the Sudakov
form factor in~\cite{Magnea:1990zb}, then applied to threshold
resummations in~\cite{Contopanagos:1997nh}. More recently, it was
shown~\cite{Magnea:2000ss} that \eq{dbeta} has the added virtue of
providing a regularization for the Landau pole, which moves off the
real axis for $\epsilon < - b_0 \alpha_s/(4 \pi)$, leading to resummed
expressions which are explicit analytic functions of the coupling and
$\epsilon$. At the level of amplitudes, a generalization of the work
of~\cite{Magnea:1990zb} to multiparton configurations has been
provided in~\cite{Sterman:2002qn}, proving an earlier
statement~\cite{Catani:1998bh} on the structure of single poles in QCD
amplitudes at two loops. This was in turn an important ingredient in
the formulation of a striking conjecture on the all-order structure of
multileg amplitudes in ${\cal N} = 4$ supersymmetric Yang-Mills
theory, for which strong evidence was provided
in~\cite{Bern:2005iz,Bern:2006vw}. A two-loop analysis of the soft
functions appearing in the exponentiation of multiparton amplitudes
was very recently started along these lines in~\cite{Aybat:2006wq}.

In the context of resummations, a key feature of the usage of the
$d$-dimensional running coupling is the fact that {\it all} infared
and collinear singularities arise by integrating the scale of the
coupling itself, while all functions appearing in resummed exponents
are finite. To illustrate how poles arise, consider the expansion of
the coupling at the three-loop level,
\beqa
  \overline{\alpha} \left( \xi^2, \alpha_s, \e \right) 
  & \hspace{-1mm} = \hspace{-1mm} &
  \alpha_s \, \xi^{- 2 \e} +
  \alpha_s^2  \, \xi^{- 4 \e} \, \frac{b_0}{4 \pi \e}
  \left( 1 - \xi^{2 \e} \right) \nonumber \\
  &&  \hspace{-2.6cm} + \,\, \alpha_s^3 \, \xi^{- 6 \e} \,
  \frac{1}{8 \pi^2 \e} \left[ \frac{b_0^2}{2 \e}
  \left( 1 - \xi^{2 \e} \right)^2 +
  b_1 \left( 1 - \xi^{4 \e} \right)
  \right],
\label{twola}
\eeqa
where $\xi$ is now a ratio of scales, while $\alpha_s$ is evaluated at
the lower (reference) scale. Clearly, \eq{twola} is finite as
$\epsilon \to 0$, but will generate poles when integrated over $\xi$.

We will now proceed to further illustrate these ideas by describing two
recent applications to resummation for processes of electroweak (EW)
annihilation, such as Drell-Yan and $Z$-boson production, or Higgs
production via gluon fusion.

\section{$N$-INDEPENDENT TERMS}

Using factorization techniques based on dimensional regularization, it
is possible to show~\cite{Eynck:2003fn} that for simple processes, such
as DIS or EW annihilation, threshold resummation can be extended, so
that all $N$-independent terms in the cross section exponentiate
together with logarithms. Consider for example the Drell-Yan cross section.
The basic step is to recognize, as suggested already in~\cite{Sterman:1987aj},
that at parton level the (collinear divergent) annihilation cross section
can be factorized as
\beq
  \omega (N, \epsilon) \, = \, \left| \Gamma(Q^2,\epsilon) \right|^2 \, 
  \psi_R(N, \epsilon)^2 \, U_R(N, \epsilon) \, ,
\label{dyp}
\eeq
where $\Gamma(Q^2,\epsilon)$ is the quark form factor, $\psi_R$ is the
real emission contribution to a special quark
distribution~\cite{Sterman:1987aj}, and similarly $U_R$ is the real
emission contribution to an eikonal function describing wide-angle
soft gluon radiation. Corrections to \eq{dyp} are suppressed by powers
of $N$. The key feature of \eq{dyp} is that real and virtual
contributions are explicitly factorized, with all virtual poles
collected in the form factor $\Gamma$. Furthermore, all functions
involved exponentiate up to $1/N$ corrections. The exponentiation of
the form factor is described in ~\cite{Magnea:1990zb}, while the
parton distribution $\psi_R$ is given by an expression of the form
\beqa
  \psi_R(N, \epsilon) & = & \exp \Bigg[ \int_0^1 dz \,
  \frac{z^{N - 1}}{1 - z} \int_z^1
  \frac{dy}{1 - y} \nonumber \\ &&  
  \kappa_\psi  \left(\overline{\alpha}
  \left( (1 - y)^2 Q^2 \right) , \epsilon \right)  \Bigg] \, .
\label{psir}
\eeqa
Similarly, the eikonal function $U_R$ can be written as
\beqa
  U_R(N, \epsilon) & = & \exp \Bigg[ - \int_0^1 dz \, \frac{z^{N - 1}}{1 - z}
  \nonumber \\ && g_U \left( \overline{\alpha} \left( (1 - z)^2 Q^2 \right),
  \epsilon \right) \Bigg] \, .
\label{ur}
\eeqa
Notice that the functions $\kappa_\psi$ and $g_U$ appearing in the
exponents have their own Feynman rules and can be independently
computed. At one loop, for example, one finds~\cite{Sterman:1987aj}
$\kappa_\psi(\alpha_s) = 2 C_F (\alpha_s/\pi) \Gamma(2 -
\epsilon)/\Gamma(2 - 2 \epsilon)$ and $g_U = - 2 C_F (\alpha_s/\pi)
\Gamma(1 - \epsilon)/\Gamma(2 - 2 \epsilon)$. In order to derive the
finite Drell-Yan partonic hard cross section, one needs to divide
\eq{dyp} by the square of a suitable quark-in-quark density. In the
\msb scheme, one may use an explicit exponential representation of the
\msb density, valid up to $1/N$ corrections, and containing only poles
generated by the running coupling . It is
\beqa
  \phi_{\msb}(N, \e) & = & \exp\Bigg[
  \int_0^{Q^2}  \frac{d \xi^2}{\xi^2} \, \Bigg\{
  B_\delta \left(\overline{\a} \left(\xi^2 \right)
  \right) \nonumber \\ & & \hspace{-1cm} \int_0^1 dz \, 
  \frac{z^{N - 1} - 1}{1 - z}
  A \left( \overline{\a} \left(\xi^2 \right) \right)
  \Bigg\} \Bigg] \, ,
\label{phims}
\eeqa
where $A(\alpha_s)$ and $B_\delta (\alpha_s)$ are, respectively, the
coefficients of the plus distribution $1/(1 - z)_+$ and of $\delta(1 -
z)$ in the Altarelli-Parisi splitting function. Clearly, \eq{phims} is
simply Ïdesigned to satisfy the Altarelli-Parisi equation and to be
constructed of poles only. A key result of ~\cite{Eynck:2003fn} is the
fact that it is possible to further factor \eq{phims} in a unique
way by isolating pure pole terms
associated with virtual contributions in a single factor. Virtual
contributions to $\phi_{\msb}(N, \e) $ take the form
\beqa
 \phi_{V} (\epsilon) \hspace{-2.1mm} & = &  \hspace{-2.1mm} \exp \Bigg\{
 \frac{1}{2} \int_0^{Q^2} \frac{d\xi^2}{\xi^2} \Bigg[
 K \left(\alpha_s, \epsilon \right) +
 \gtil \left(\overline{\alpha} (\xi^2) \right) \nonumber \\  && + \,
 \frac{1}{2} \int_{\xi^2}^{\mu^2} \frac{d \lambda^2}{\lambda^2}~
 \gamma_K \left( \overline{\alpha} (\lambda^2) \right) 
 \Bigg] \Bigg\},
\label{phiv}
\eeqa
which mimicks the exponentation of the Sudakov form factor. In fact,
$\gamma_K (\alpha_s)$ and $K(\alpha_s, \epsilon)$ are, respectively,
the cusp anomalous dimension and the counterterm function featuring in
the form factor resummation, while $\gtil (\alpha_s)$, independent of
$\epsilon$, is constructed recursively from the analogous function
present in $\Gamma(Q^2, \epsilon)$.  $\phi_{V} (\epsilon)$ is designed
to cancel exactly the virtual poles associated with the form factor.
As a consequence, the Drell-Yan hard part $ \widehat{\omega} (N) \equiv
\omega (N, \epsilon)/(\phi_{\msb}(N, \e))^2$ can be written as the
product of virtual and real contributions, which are separately
finite.  They are
\beqa
  \widehat{\omega}_V (Q^2) & = & \frac{\left| \Gamma (Q^2, \epsilon) 
  \right|^2}{(\phi_V(\epsilon))^2}~,
  \nonumber \\
  \widehat{\omega}_R (N) & = & \left[ U_R (N, \epsilon) 
  \left( \frac{\psi_R(N, \epsilon)}{\phi_R(N, \epsilon)}
  \right)^2  \right]~,
\label{hatdy}
\eeqa
where $\phi_R$ is simply defined as $\phi_{\msb}/\phi_V$. Taking the
limit $\epsilon \to 0$, one recovers the usual resummation formula for
the Drell-Yan cross section in the \msb scheme, where however all
$N$-independent terms are now exponentiating, along with threshold
logarithms. The final expression is
\beqa
  \widehat{\omega}_{\msb}(N) & = & \left|\frac{\Gamma(Q^2,
  \epsilon)}{\phi_V(Q^2,\epsilon)} \right|^2
  \exp \Bigg[ \, F_{\msbns} (\alpha_s) \nonumber \\ && \hspace{-1,2cm} + \, 
  \int_0^1 \! dz \, \frac{z^{N - 1} - 1}{1 - z} 
  \Bigg\{ D \left(\alpha_s \left((1 - z)^2 Q^2 \right) \right) 
  \nonumber \\ && \hspace{-1.2cm} + \, \, 2 \, \int_{Q^2}^{(1 - z)^2 Q^2}
  \frac{d \mu^2}{\mu^2} \, A \left(\alpha_s(\mu^2) \right)
  \Bigg\} \Bigg] \, ,
\label{long}
\eeqa
where the limit $\epsilon \to 0$ is understood in the virtual
terms. Analogous formulas hold for the DIS cross section, for the
Drell-Yan cross section in the DIS scheme, and for Higgs production
via gluon fusion (where the gluon form factor replaces the quark form
factor). Clearly, the exponentiation of $N$-independent terms does not
have the predictive power of the resummation of towers of logarithms,
since, for example, the function $F_{\msbns}(\alpha_s)$ receives novel
contributions at every order.  It should be emphasized, however, that
$F_{\msbns}$ has a precise definition of its own, and can be computed
in principle without resorting to finite order results for the cross
section. Furthermore, a large fraction of constant terms arising in
ordinary perturbation theory are actually generated through
exponentiation at lower orders, and, as we will see shortly, having
defined $F_{\msbns}$ helps uncover new structure to all orders.

\section{EW ANNIHILATION AND DIS}

Our second application is the derivation of a general relation
expressing the coefficients of the resummation for EW annihilation
processes in terms of data computed in
DIS~\cite{Laenen:2005uz}. Within this formalism, the result can be
extracted making use of the finiteness of equations like \eq{hatdy},
which follows from factorization. Consider specifically the real part
of the cross section, $\hat{\omega}_R(N)$. While the numerator
contains genuine information associated with Drell-Yan kinematics, the
denominator is completely determined by knowledge which can be
extracted from DIS, specifically form factor and splitting function
information. Imposing the cancellation of all infrared and collinear
poles yields relations between resummation coefficients for the two
processes. It is straightforward to evaluate all functions involved at
one loop: at this level, taking the limit $\epsilon \to 0$ and
comparing the result with the real contribution to \eq{long} one finds
immediately
\beq
 D^{(1)} \, = \, 4 B_\delta^{(1)} - 2 \gtil^{(1)} \, = \, 0 \, ,
\label{d1}
\eeq
a well-known result.

At two loops, the relevant DIS information can be extracted
from~\cite{Herrod:1980rm,Curci:1980uw}, while the second order
coefficient of the function $\gtil (\alpha_s)$ was computed
in~\cite{Eynck:2003fn}. The finiteness of \eq{hatdy} leads to
\beqa
  D^{(2)} & = & 4 B_\delta^{(2)} - 2 \gtil^{(2)} - 
  \frac{b_0}{2} F_{\msb}^{(1)} \nonumber \\
  & = & \left(- \frac{101}{27} + \frac{11}{3} \zeta (2) + \frac{7}{2}
  \zeta (3) \right) C_A C_F \nonumber \\
  & & + \, \left(\frac{14}{27} - \frac{2}{3} \zeta (2)
  \right) n_f C_F \, .
\label{d2}
\eeqa
This result was previously obtained
in~\cite{Contopanagos:1997nh,Vogt:2000ci}, by comparing the results of
resummation with the two-loop calculation of
Ref.~\cite{Hamberg:1991np}, along the lines of~\cite{Magnea:1991qg}.
One can, finally, push the calculation to the three-loop level, thanks
to the remarkable results of~\cite{Moch:2004pa,Moch:2005id}, where the
three-loop nonsinglet splitting functions and form factor were
explicitly computed. The result is
\beqa
  D^{(3)} & = & 4 B_\delta^{(3)} - 2 \gtil^{(3)} - b_0 F_{\msb}^{(2)} -
  \frac{b_1}{2} F_{\msb}^{(1)} \nonumber \\  & = &  
  \left(- \frac{297029}{23328} +
  \frac{6139}{324} \zeta (2) -
  \frac{187}{60} \zeta^2 (2) \right. 
  \nonumber \\  & & \left. \hspace{-1,1cm} + \, \frac{2509}{108} \zeta (3) -
  \frac{11}{6} \zeta (2) \zeta(3) - 6 \zeta(5) \right) C_A^2 C_F
  \nonumber \\ & & \hspace{-1.1cm} + \,
  \left(\frac{31313}{11664} - \frac{1837}{324} \zeta (2) \right. 
  \label{d3} \\ & & \left. \hspace{-1.1cm} + \,\, \frac{23}{30}
  \zeta^2 (2) - \frac{155}{36} \zeta (3) \right) n_f C_A C_F
  \nonumber \\ & & \hspace{-1.1cm} + \,
  \left(\frac{1711}{864} - \frac{1}{2} \zeta (2) -  \frac{1}{5}
  \zeta^2 (2) - \frac{19}{18} \zeta (3) \right) n_f C_F^2
  \nonumber \\ & & \hspace{-1.1cm} + \,
  \left(- \frac{58}{729} + \frac{10}{27} \zeta (2) + \frac{5}{27}
  \zeta (3) \right) n_f^2 C_F \, . \nonumber
\eeqa
The coefficient $D^{(3)}$ was computed simultaneously and
independently by~\cite{Moch:2005ky}, and the result later checked with
different methods in~\cite{Idilbi:2005ni,Ravindran:2005vv}.

The expressions we have derived for the perturbative coefficients of
the function $D(\alpha_s)$ up to three loops, in terms of
$B_\delta(\alpha_s)$, $\gtil(\alpha_s)$ and $F_{\msb}(\alpha_s)$ are
strongly suggestive of a simple all-order relation. Since it is
well-known that the function $A(\alpha_s)$ coincides with (one half
of) the cusp anomalous dimension $\gamma_K(\alpha_s)$, one finds that
in fact threshold resummation for the Drell-Yan process, at $g$ loops
in the exponent, is completely determined dy DIS data at $g$ loops,
plus the knowledge of $N$-independent terms for Drell-Yan at $g - 1$
loops. The all-order results are
\beqa
  A (\alpha_s) & = & \gamma_K (\alpha_s)/2 ~, \nonumber  \\
  D (\alpha_s) & = & 4 \, B_\delta (\alpha_s) - 2 \, \gtil (\alpha_s)
  \nonumber \\ & & + \, \hat{\b} (\alpha_s) \, \frac{d}{d \alpha_s} 
  F_{\msb} (\alpha_s) \, .
\label{allo}
\eeqa
A formally identical relation ties together the gluon annihilation
cross section with the singular terms in the gluon splitting function
and with the gluon form factor. Remarkably, up to three loops, the
perturbative coefficients of the functions $A$ and $D$ in the two
cases differ only by the replacement of the overall factor of $C_F$
for quarks with a factor $C_A$ for gluons. In other words, to this
order the resummation is only sensitive to the color representation of
the Wilson lines replacing the hard annihilating partons in the soft
approximation. This simple replacement rule, however, is not expected
to hold at yet higher orders.

\section{PERSPECTIVE}

We have given a short review of some of the results that can be
obtained tackling threshold resummation with the tools provided by
factorization and dimensional regularization. From
a phenomenological viewpoint, perhaps the most interesting result is
the calculation of $D^{(3)}$, which contributes to resummation for EW
annihilation at the ${\rm N^3LL}$ level. In fact, the only missing
contribution in order to resum exactly to that accuracy is the
four-loop cusp anomalous dimension $\gamma_K^{(4)}$, which in
principle lies close the current boundaries of computability. In any
case it can be convincingly shown~\cite{Moch:2005ba} that
$\gamma_K^{(4)}$ makes a numerically negligible contribution to the
cross section. Having at our disposal, with a good approximation, four
towers of logarithms for both DIS and EW annihilation, we can
stringently test the level of convergence of the perturbative
expansion, both with and without resummation. Preliminary tests
show~\cite{Moch:2005ky,Moch:2005ba} that resummed perturbation theory
converges well across much of the kinematical range relevant for
Tevatron and the LHC.

\end{document}